\documentclass[trackchanges]{aastex62} % ,trackchanges
\usepackage{amsmath,amstext}

\newcommand{\okina}{\textquoteleft}

\newcommand{\Ou}{{\okina}Oumuamua}
\newcommand{\be}{\begin{eqnarray}}
\newcommand{\ee}{\end{eqnarray}}
\usepackage{xcolor}
\usepackage{comment}

\newcommand{\MSun} {\mbox{$M_{\odot}$}}

\received{\today}
\submitjournal{ApJ}

\shorttitle{The fate of interstellar objects during cloud collapse}
\shortauthors{Pfalzner, Paterson, Bannister \& Portegies Zwart}
\begin{document}

\title{Interstellar objects follow the collapse of molecular clouds}

\author[0000-0002-5003-4714]{Susanne Pfalzner} 
\affiliation{J\"ulich Supercomputing Center, Forschungszentrum J\"ulich, 52428 J\"ulich, Germany}
\affiliation{Max-Planck-Institut f\"ur Radioastronomie, Auf dem H\"ugel 69, 53121 Bonn, Germany}

\author[0000-0002-6652-0303]{Dylan Paterson}
\affiliation{School of Physical and Chemical Sciences | Te Kura Mat\={u}, University of Canterbury,
Private Bag 4800, Christchurch 8140,
New Zealand}

\author[0000-0003-3257-4490]{Michele T. Bannister}
\affiliation{School of Physical and Chemical Sciences | Te Kura Mat\={u}, University of Canterbury,
Private Bag 4800, Christchurch 8140,
New Zealand}

\author[0000-0002-5003-4714]{Simon Portegies Zwart} 
\affiliation{Leiden Observatory, Leiden University, PO Box 9513, 2300 RA
Leiden, the Netherlands}

\email{s.pfalzner@fz-juelich.de}

% 250 word limit, this is good
\begin{abstract}
Interstellar objects (ISOs), the parent population of 1I/\Ou\ and 2I/Borisov, are abundant in the interstellar medium of the Milky Way. This means that the interstellar medium, including molecular cloud regions, has three components: gas, dust, and ISOs. From the observational constraints for the field density of ISOs drifting in the solar neighbourhood, we infer a typical molecular cloud of 10 pc diameter contains some 
10$^{18}$ ISOs.  At typical sizes ranging from hundreds of metres to tens of km, ISOs are entirely decoupled from the gas dynamics in these molecular clouds. Here we address the question of whether ISOs can follow the collapse of molecular clouds. We perform low-resolution simulations of the collapse of molecular clouds containing initially static ISO populations toward the point where stars form. In this proof-of-principle study, we find that the interstellar objects definitely follow the collapse of the gas --- and many become bound to the new-forming numerical approximations to future stars (sinks). 
At minimum, 40\% of all sinks have one or more ISO test particles gravitationally bound to them for the initial ISO distributions tested here. This value corresponds to at least $10^{10}$ actual interstellar objects being bound after three initial free-fall times.
Thus, ISOs are a relevant component of star formation. 
We find that more massive sinks bind disproportionately large fractions of the initial ISO population, implying competitive capture of ISOs. 
Sinks can also be solitary, as their ISOs can become unbound again --- particularly if sinks are ejected from the system.
Emerging planetary systems will thus develop in remarkably varied environments, ranging from solitary to richly populated with bound ISOs. 
\end{abstract}


\section{Introduction} 
\label{sec:intro}

We are now in an era where the Galaxy's interstellar medium (ISM) is known to have a third component. In addition to gas and dust, the ISM contains interstellar objects (ISOs) --- a distribution of free-drifting minor planets and smaller objects. 
Unlike the dust in the ISM, the ISOs are decoupled from the gas due to their larger size. 
Here we address whether ISOs nevertheless take part in molecular cloud collapse due to the gravitational forces alone.\\ 
Although the existence of ISOs was first hypothesised nearly half a century ago \citep{Whipple:1975, Sekanina:1976}, 
confirmation came only in the last four years, with the discovery of 1I/\Ou\
\citep{Meech:2017} and 2I/Borisov\footnote{Initially designated C/2019 Q4 (Borisov) in MPEC 2019-R106 \url{https://www.minorplanetcenter.net/mpec/K19/K19RA6.html} \citep{Borisov:2019}.}
The observed constraints on the visible volume of space within the Solar System over a 17-year period, which includes the detection of 1I, imply that a spatial number density of 10$^{15}$ ISOs 
exists per cubic parsec in the solar neighbourhood \citep{Engelhardt:2017,Meech:2017,PortegiesZwart:2018}.
Assuming that this is a characteristic density across the entire Milky Way, vast populations of ISOs must drift in our Galaxy \citep{MoroMartin:2009,ISSI:2019}. 
While gas and dust are often described as being the primary components of the ISM \citep{Klessen:2016S}, this new understanding demands that we have to take minor planets into account as a key ISM component \citep{Pfalzner:2020}. 

Gas and dust are unevenly spread in the interstellar medium, concentrating at densities of $\rho <10^2$ particles/cm$^3$ in molecular clouds (MCs), which are the birthplaces of stars. 
Clouds typically have sizes of 5 to 200 pc, masses of 10$^3$ to 10$^7$ \MSun\ and a temperature around 10 K \citep{Murray:2011}.
In spiral galaxies, such as the Milky Way, they are primarily located in the disk.
Star formation appears to take place exclusively within MCs, and only when they become unstable and start to collapse under their own weight \citep{Bergin:2007, Kennicutt:2012}. 
In the collapse, clumps of high-density gas form, which later develop into protostars, and eventually become stars. 
Crucially, the dust follows the general collapse of the gas, as the dust is coupled to the gas movement. The forming stars are often surrounded by gas-dust disks that can develop into planetary systems; it is only the presence of the dust that allows the formation of planets.

In a molecular cloud collapse, it is rare for just a single star to be created: mostly, an entire group emerges \citep{Lada:2003,Porras:2003,Pfalzner:2009,Portegies:2010}. 
The formation of a cluster of stars during cloud collapse has been studied extensively with computer simulations \citep{Bate:2003, Bonnell:2003MNRAS.343..413B, Tan:2006,vazques:2017, Riaz:2018, Lee:2019, Verliat:2020, Dobbs:2020, Wall:2020}. 
During the initial collapse phase in molecular clouds, the gas density eventually leads to the formation of stars. The dust density obeys the same flow of movement as the gas, as the $\sim$ 100 $\mu$m particles of dust are coupled to the gas \citep{Testi:2014}. The coupling strength in the Epstein regime \citep{Epstein:1924,Weidenschilling:1977} depends to the first order on gas density and the particle size, with the strength of the dust-gas coupling weakening for larger particles. As ISOs are some six to eight orders of magnitude larger than the dust (cf. 1I at an effective radius of 49-220 m \citep{Trilling:2018} and 2I at a nuclei radius of 0.2--0.4~km \citep{Jewitt:2019b,Hui:2020}), the ISOs are completely decoupled from the cloud's gas.

From the inferred density of interstellar objects throughout the interstellar medium, a typical 10-pc-diameter molecular cloud should contain $\gtrapprox 10^{18}$ ISOs; large MCs would hold $\sim 10^{21}$. 
Many of these ISOs may just be passing through. 
In the solar neighbourhood, ISOs spend on the order of 0.1--0.2\% of their time passing through MCs \citep{Pfalzner:2020}.
For slow ($<$ 10 km/s) ISOs that are closer to the centre of the Galaxy ($\sim$ 4.5 kpc from the Galactic centre), this can increase to several per cent of their time. 
It is unknown whether ISOs start off denser in MCs relative to their field density: during the formation of molecular clouds, the ISO density could be enhanced together with that of the gas and dust.
The current observational constraint from the Solar System only gives a density outside of a molecular cloud. 
Alternatively, fully-formed MCs may weakly capture slow ISOs, which could further increase the population in clouds \citep{Pfalzner:2019}. 

Here we investigate what happens to the ISOs in a molecular cloud during its collapse. 
Our central goal is to test whether the ISOs follow the gravitational potential of the gas when collapse sets in. We find that gravitational forces alone are sufficient for this to happen (Sec.~\ref{sec:results}): the ISOs not only follow the gas but become bound to the star formation sub-clump. In Sec.~\ref{sec:method} we describe the numerical method, which is similar to those mentioned above, with the novel difference of including a population of test particles representing the ISO population. As we intend here only a qualitative proof-of-concept, these simulations are low-resolution, neglecting effects such as stellar feedback. Nevertheless, they suffice to determine the parameters that are relevant in this novel context. In section \ref{sec:discussion}, we detail the challenging requirements for a statistically relevant quantitative assessment. We provide a first impression on the degree to which the ISO density follows the gas density, the numerical stellar-precursor sink-formation in the hydrodynamics, and eventually the ISOs' trajectories around these sinks. In Sec.~\ref{sec:discussion} we discuss the limitations of this investigation, and the next steps to be taken in future studies. Finally, we explore the implications of bound populations of ISOs being present during the star-formation process.\\

\section{Method}
\label{sec:method}

% Table 1.
\begin{table}[b]
%\centering
\caption{Simulation parameters}
\begin{tabular}[t]{l|rrlrrrr}
\tablewidth{0pt}
\hline 
\hline
	model  &  $N_{\text{gas}}$ & $N_{\text{ISO}}$ & ISO distribution  &  $
	[\MSun ]$ &  $R [$pc$]$ & $\tau_{\text{ff}}[Myr]$  \\
 \hline
Pl-30k  &  $30\,923$  &  $30\,000$ & Plummer & 3000  & 3  &   0.76 \\
Pl-60k &  $62\,079$  & $60\,000$  & Plummer &  3000  & 3  &   0.76  \\
Fr-30k & $30\,923$   &  $30\,000$ & fractal  & 3000   & 3  &  0.76  \\
Fr-60k & $62\,079$   & $60\,000$  & fractal  & 3000   & 3  & 0.76   \\ 
\hline 
\end{tabular}
	\vspace{0.5em}
	\tablecomments{Col.~1 gives the model name, $N_{\text{gas}}$ is the number of test particles for the gas in the model, $N_{\text{ISO}}$ the number of ISO test particles, Col.~4 the initial distribution of the ISOs, Col.~5 the mass of the molecular cloud $r_{\text{hm}}$, Col.~6 the radius of the simulation box, and $\tau_{\mbox{ff}}$ the corresponding free-fall time.}
	\label{tab:set-up}
      \end{table}
      
We simulate the dynamical evolution of interstellar objects (ISOs) within a collapsing molecular cloud.
These calculations start from a distribution of gas that becomes gravitationally unstable, together with a population of test particles representing the ISOs. 
The ISOs are decoupled from the hydrodynamics of the gas, but they feel its gravitational potential. 
We follow the cloud's gravitational collapse using the smoothed particle hydrodynamics (SPH) method \citep{Monaghan:2005}, examining the system's evolution for five initial free-fall time scales. We adopt a very simplified approach for the description of the cloud collapse and the resulting star formation process. This is
motivated by the established complexity of the physics, their implementation details, and the sensitivity of such results on the initial conditions \citep{10.1093/mnras/stw781}. 
Here we are interested in the key question: whether ISOs do follow any of the cloud collapse process.
Once this is established, its dependency on the details of the star formation process can be addressed by more advanced star formation realisations.

The simulations are performed within the Astrophysics Multi-purpose Software Environment (AMUSE; \cite{PortegiesZwart2013456,2018araa.book.....P}), which allows different codes to be easily combined to efficiently simulate the hydrodynamics and gravitational forces. 
In this work, the SPH code {\tt Fi} \citep{1989ApJS...70..419H, 2006ApJ...645.1024P} is used to simulate the hydrodynamics. 
We adopted a non-isothermal equation of state without integrating the entropy \citep[using a polytropic gas index of $\gamma =1$][]{1989ApJS...70..419H} but use the thermal-model cooling as discussed in \citet{2012MNRAS.420.1503P}\footnote{See also chapter 5 (page 5-35) of \cite{2018araa.book.....P}.}The gas cooling and the hydrodynamics were interlaced using the bridge method discussed in \cite{2018araa.book.....P}. The gas cooling routine adopted half the timestep of the hydrodynamics solver.
We used an artificial viscosity of $\alpha = 0.01$, and an SPH smoothing kernel of 0.4\,pc, which results in a timestep of almost 5000 years. 
The timestep in the calculation, however, is dynamically determined dependent on local density and temperature. 
In our simulations, the hydrodynamics timestep varies from 20 to 2000\,yr.

We investigate a spherical initial gas distribution for the molecular cloud, modelled with $6 \times 10^4$ equal-mass gas particles (for simulation properties, see Table \ref{tab:set-up}). 
The gas is initialized as an isotropic homogeneous sphere, with a divergence-free random Gaussian velocity field $\delta \Vec{v}$ with a power spectrum $|\delta v|^2\propto k^{-3}$ from wavenumber $k=2$ to $k=32$ \citep{2003MNRAS.343..413B}. Here $k = 2 \pi/D$ where $D$ is the diameter of the cloud \citep[for more details, we refer to][]{2020arXiv200309011W}.
The initial radius of the gas distribution is 3\,pc and the total initial cloud mass in each simulation is 3000\,\MSun. 
While this cloud size and mass are on the low side of the spectrum of observed cloud properties \citep{2004A&A...414..633G}, we aim here at a proof-of-principle rather than an exhaustive cloud parameter exploration. 
Such a cloud will collapse, due to the local low-pressure regions in the turbulent field.
It starts to form high-density clumps in about a single initial free-fall time scale $\tau_{\mbox{ff}}$, which corresponds to 0.76\,Myr for these parameters.

As the cloud collapses, we use the common numerical mechanism of replacing the peaks of high gas density by sink particles, as described in \citet{Federrath_Sink_Particles}.
Simulations of the cloud collapse become computationally expensive when the density contrast increases during the cloud collapse. 
At this point, it becomes too computationally costly to resolve the high-density regions where the stars are forming.  
Sink particles are instantiated to replace the peaks of gas density once the local gas density exceeds a mean-local density of $8.5\times 10^{-22}$\,g/cm$^3$, as discussed in 
\cite{1997A&A...325..972G}.
A sink particle is spherical, with a sink radius of $0.05$\,pc, and it inherits the mass, position and velocity of the SPH particle.
The sink particles merge as soon as two sinks approach either within a sink radius, forming a new sink with the total mass in the centre of mass of the two sinks. 
During the merger, linear momentum is preserved.
Sink particles are often regarded as representative for one or more stars \citep[cf.][]{Federrath_Sink_Particles}: this is the case in our simulations.

The gas particles of the molecular cloud interact both hydrodynamically and through their self-gravity. As the number of gas particles is rather large for a direct summation approach at $6 \times 10^4$, the Barnes-Hut tree code \citep[BHTree;][]{BARNES:1986} is used for modelling the self-gravity of the gas \citep{Pfalzner:1996,2014hpcn.conf...54B}. 
In contrast, as only a few dozen sink particles are created in these simulations, their self-gravity is calculated to machine precision using direct N-body integration, for which we use {\tt Huayno} \citep{2014A&A...570A..20J}.

We augment this standard method for investigating clustered star formation by adding an additional $6 \times 10^4$ test particles to model the ISO population. 
These massless ISO test particles experience acceleration due to the gravitational potential of the gas and sink particles, but do not themselves contribute to the potential. 
The ISOs were integrated as test particles with the coupled-component integrator in the {\tt Huayno} solver.
In our simulations here, we consider the case where the ISOs are not in relative movement to the cloud.
Low relative velocities may be highly probable if the majority of ISOs are from young systems \citep{Meech:2017,PortegiesZwart:2018,ISSI:2019}.
A broad relative velocity distribution has long been posited for the overall Galactic ISO population based on their origins from stellar systems \citep{McGlynn:1989}; we defer this more complex situation to future work.
The ISO test particles are initially distributed within the same 3\,pc radius as the gas. 
To generate the initial conditions for the ISO positions and velocities relative to the centre of mass, a different random seed was adopted each time.
If we take the inferred ISO density of 10$^{15}$\,pc$^{-1}$ in the interstellar medium as guidance, such an MC would typically contain on the order of 10$^{15}$--- 10$^{18}$ ISOs. 
This number of particles is currently beyond reach for such a simulation; thus, the resolution is lower, with each test particle here standing in for roughly $10^{10}$ ISOs.

\begin{figure}[t]
\scalebox{.57}{
\plotone{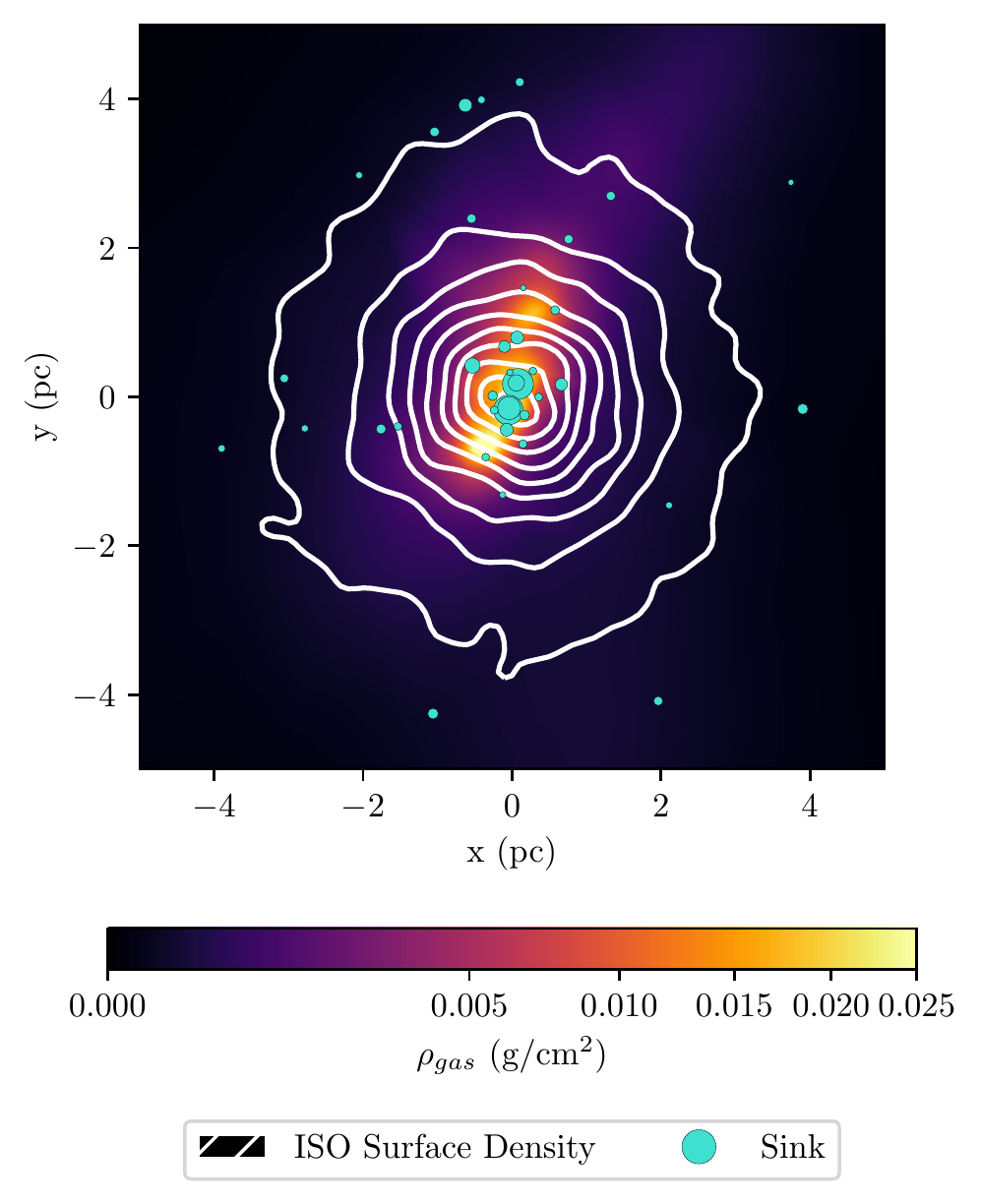}
\plotone{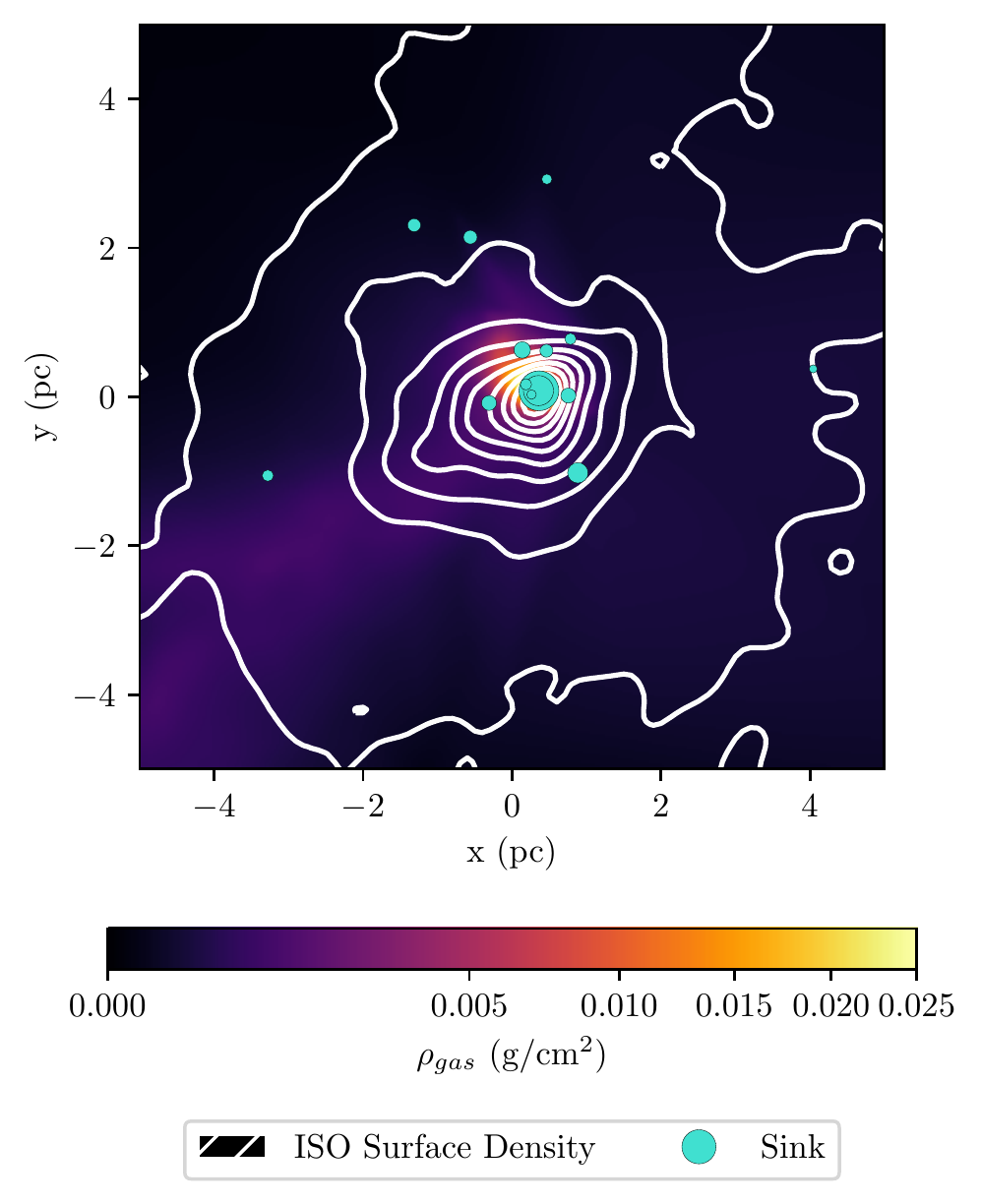}}
\caption{
Snapshots of two of the four simulations of the gravitational collapse of a molecular cloud that include interstellar objects.
The full movies of all four simulations are available at \doi{10.5281/zenodo.4301100}.
{\bf Left:} molecular cloud with $6\times10^4$ ISO test particles initially distributed as a Plummer sphere with a virial radius of 3\,pc.
{\bf Right:} fractal cloud with $6\times10^4$ ISO test particles initially distributed with a fractal dimension of 1.6. 
In both, the background colour indicates the column density  distribution of the $6 \times 10^4$ gas particles in g/cm$^3$.
By three free-fall times, a similar number of sink particles have formed from both gas distributions (turquoise dots, in projection; sink mass scales to size of dot).
The surface density of interstellar objects is tracked by $6 \times 10^4$ initial test particles (white contours; in projection).
}
\label{fig:impression}
\end{figure}

Currently, the spatial distribution of ISOs within molecular clouds is unknown. 
Here we look at two reasonable models for the initial distribution of the ISO particles. 
First, a fractal distribution \citep{2004A&A...413..929G} with a fractal dimension of 1.6, distributed according to the fractal potential, as clustered star formation often shows a filamentary structure \citep[][]{Andre:2014}. 
Second, we investigate a Plummer sphere distribution \citep{1911MNRAS..71..460P}, to explore a situation where the ISOs of the ISM take part in the formation process of a molecular cloud, and become more concentrated in the centre of the molecular cloud.
The initial fractal distribution for the ISOs is generated using the AMUSE routine {\tt new\_fractal\_cluster\_model()}, and for the Plummer distribution with
 \be
\rho_P = \left( \frac{3M}{4 \pi a^3} \right)\left( 1 +\frac{r^2}{a^2} \right)^{-5/2},
\ee
where $M$ is the total cluster mass, $a$ a scale parameter for the cluster core size and $r$ the radial distance to the centre of the potential distribution. 
For both types of ISO distributions, we performed simulations, each with $3\times10^4$ and $6\times10^4$ particles, to get a first impression of the effect of resolution on the results. These are referred to throughout this work as Fr-30k and Fr-60k for the fractal distributions and Pl-30k and Pl-60k for the Plummer-type simulations (see Table~\ref{tab:set-up}).

\section{Results}
\label{sec:results}

\begin{figure}[t]
\includegraphics[width=0.49\textwidth]{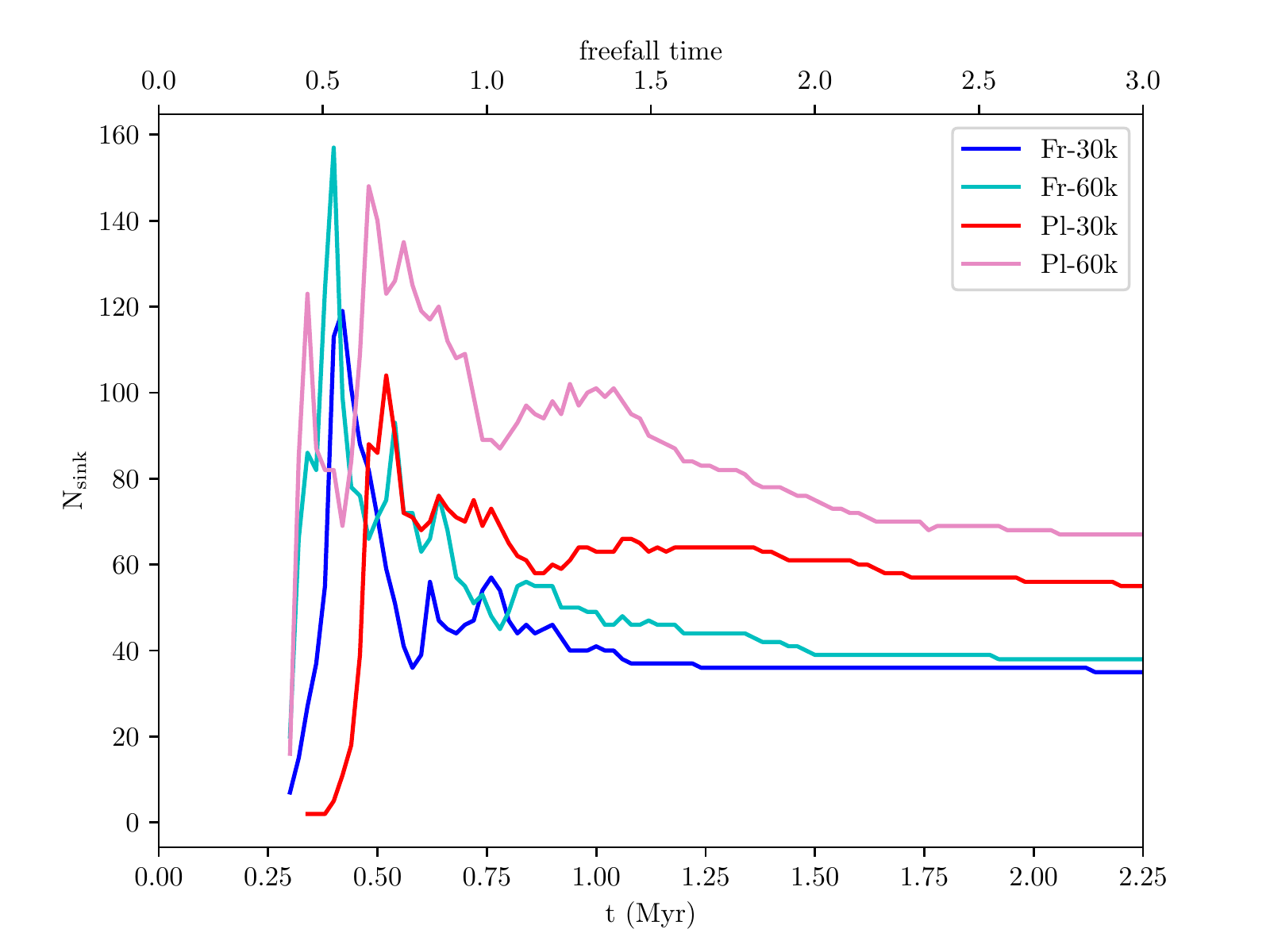}
\includegraphics[width=0.49\textwidth]{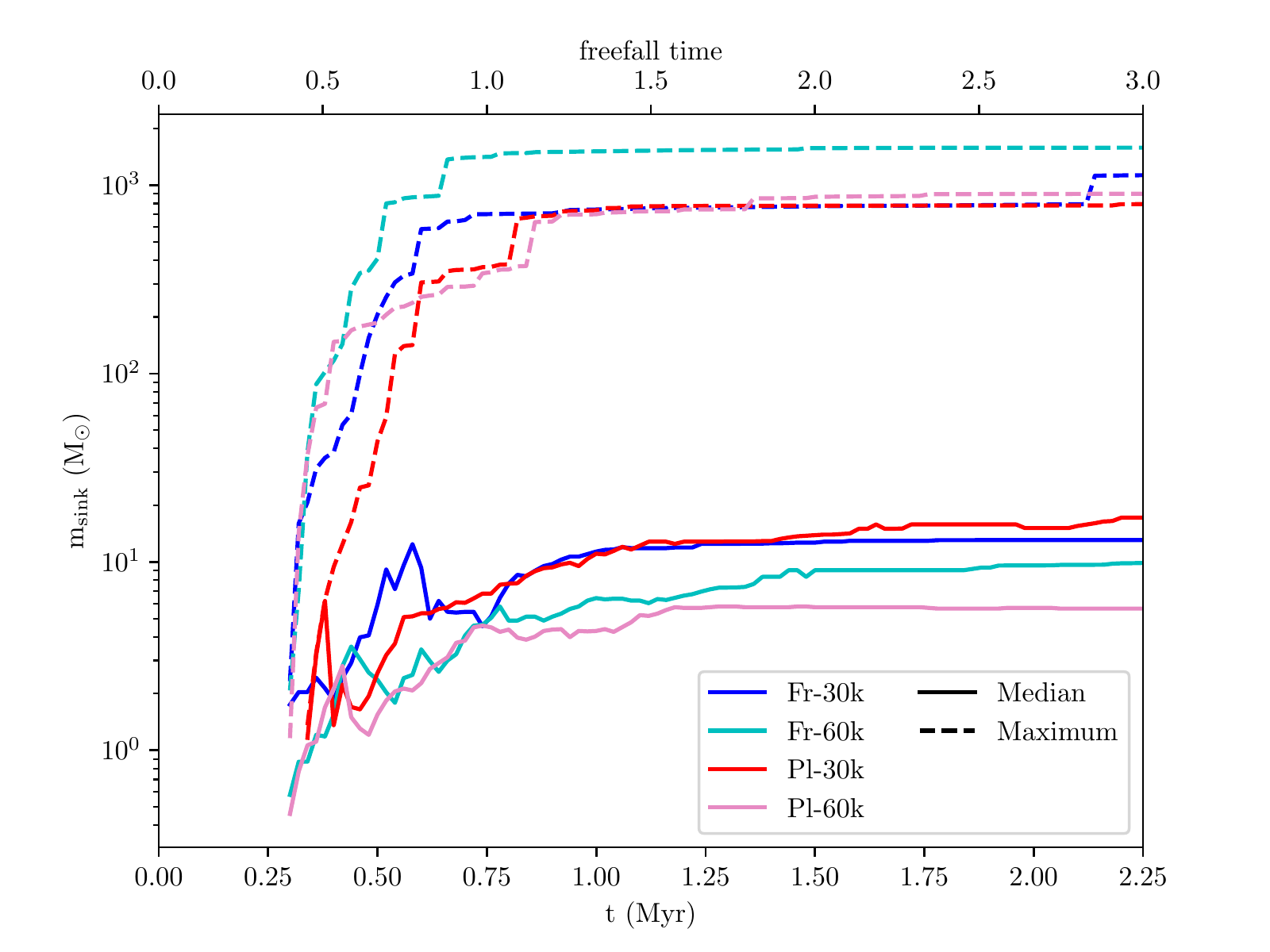}
\caption{
Formation and mass evolution of the sink particles in all four molecular cloud collapse simulations, from formation from the initial gas distribution through to the physically plausible simulation limit of three free-fall times.
{\bf Left:} number of discrete sinks with time. {\bf Right:} the mass bound into sinks as a function of time, with population variation shown through both the median and maximum sink masses. 
Simulations are distinguished by line colour: ``Fr'' is an initial fractal distribution of gas with dimension 1.6,  ``Pl" is an initial Plummer distribution of gas, while 30k and 60k indicate the number of gas particles used in the hydro code. 
}
\label{fig:sinkproperties}
\end{figure}

We assess the outcomes of our four simulations of the gravitational collapse of a molecular cloud containing interstellar objects.
Their dynamical development is illustrated via movies\footnote{The full animated versions are at \doi{10.5281/zenodo.4301100}.} of the gas density and ISO surface density.  

The Fig.~\ref{fig:impression} snapshots show that in both Pl-60k and Fr-60k, several tens of star-forming pockets --- the sink particles --- have formed from the respective gas distributions at three initial free-fall times. 
From these sinks, a wide range of stellar systems could eventually form: single stars, binaries, or even small clusters of stars. 
At the three free-fall time endpoint of our simulations, the sink masses are typically between 1 and 30 \MSun. 
These high sink masses are a direct result of the relatively low resolution of our simulations, as is the high star formation efficiency.
Nevertheless, such low resolutions should suffice for a proof-of-principle investigation. 
Many of the sink particles concentrate around the simulation space's center, as one would expect from a forming star cluster. 
However, the sink dynamics are complex, strong and forceful: throughout the simulations, sinks grow, merge, and some are even ejected from the forming cluster of stars. 

We first assess how the local density of ISOs relates to the gas distribution and the sinks.
We determine the local density of ISOs in the simulations, $\rho_{\mbox{iso}}$, by measuring the distance to the tenth-nearest neighbor for each sink as
\be \rho_{\mbox{iso}}=  \frac{n-1}{\pi r_n^2},\ee 
where $r_n$ is the distance to the n-th neighbour. The number of neighbours $n$ has to be chosen considering that high values of $n$ give a lower spatial resolution, but smaller fractional uncertainty \citep{Casertano:1985}. Here we chose $n$=10.
The ISO surface density is then obtained by summing in perspective over the local ISO densities; the white contours indicate the result. 
In Fig.\,\ref{fig:impression} and the corresponding movies, the background color gives the gas's surface density distribution in g/cm$^3$, as indicated by the color bar. 
The interstellar objects' density is centred around the central stellar cluster: the ISOs follow the gas density and begin to orbit around the forming sinks. 
This perspective on the simulations suggests that ISOs are more subsequently concentrated when initialised in the Plummer profile than for the fractal distribution.

\begin{figure}[t]
\includegraphics[width=0.48\textwidth]{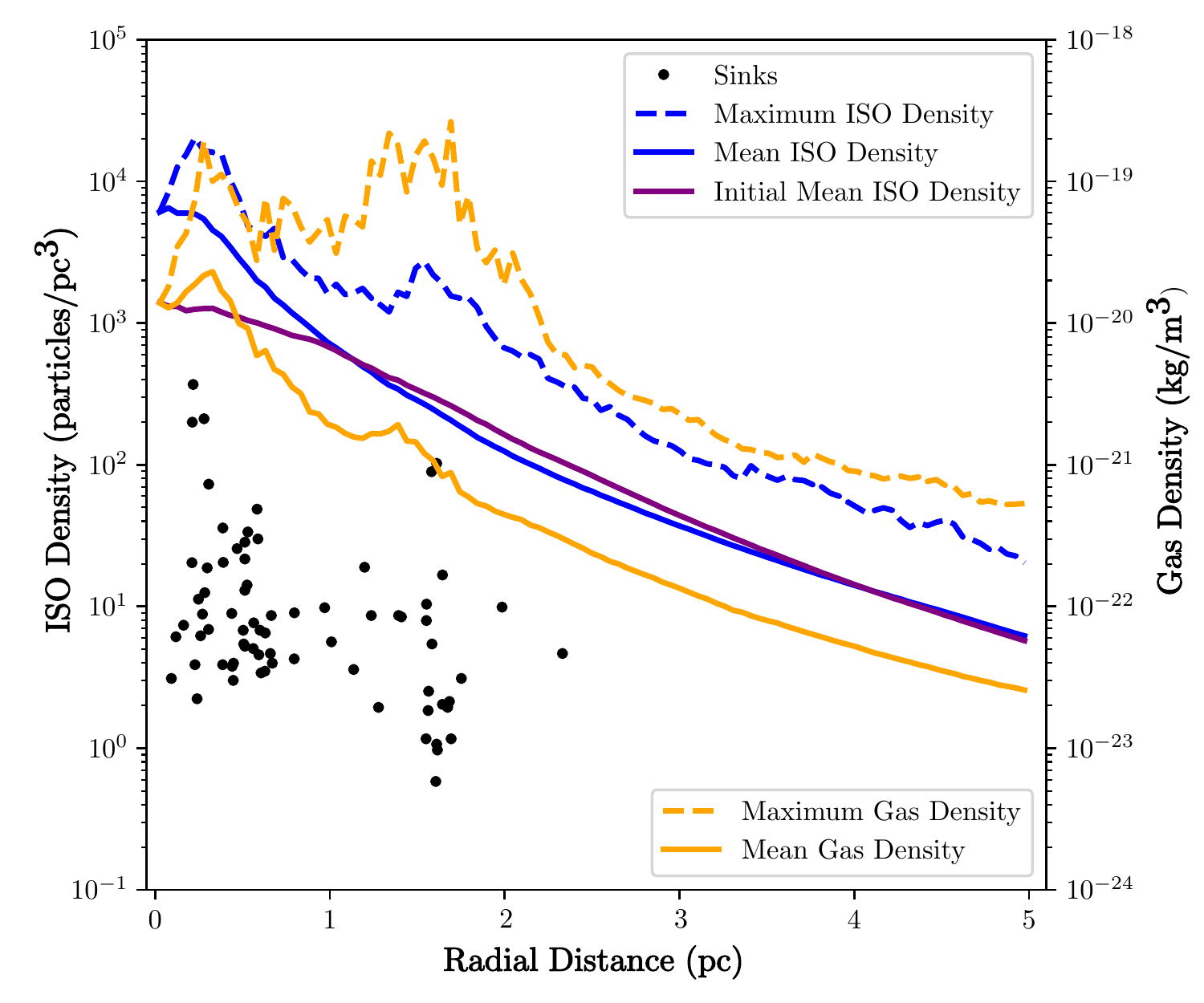}
\includegraphics[width=0.48\textwidth]{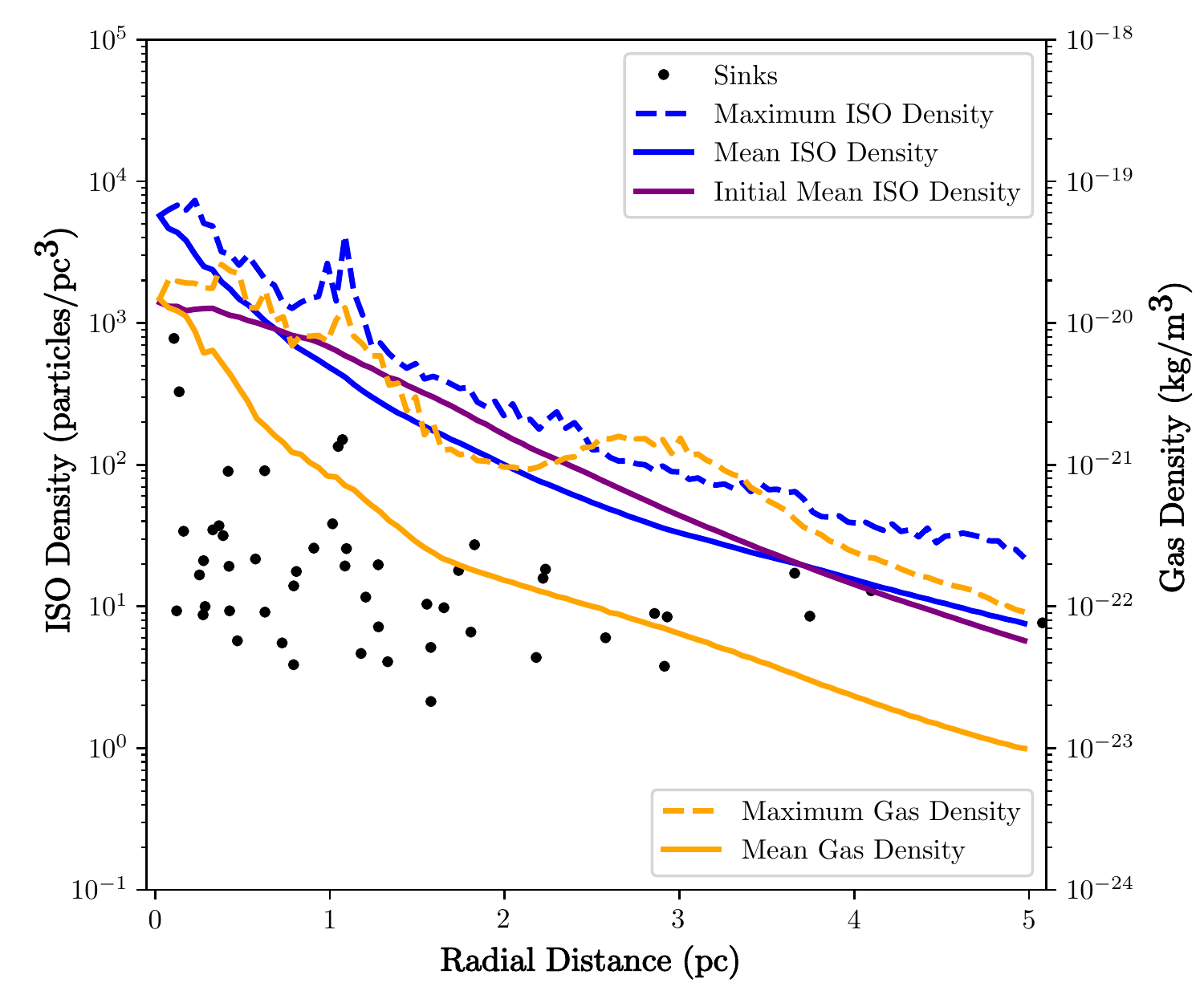}
\includegraphics[width=0.48\textwidth]{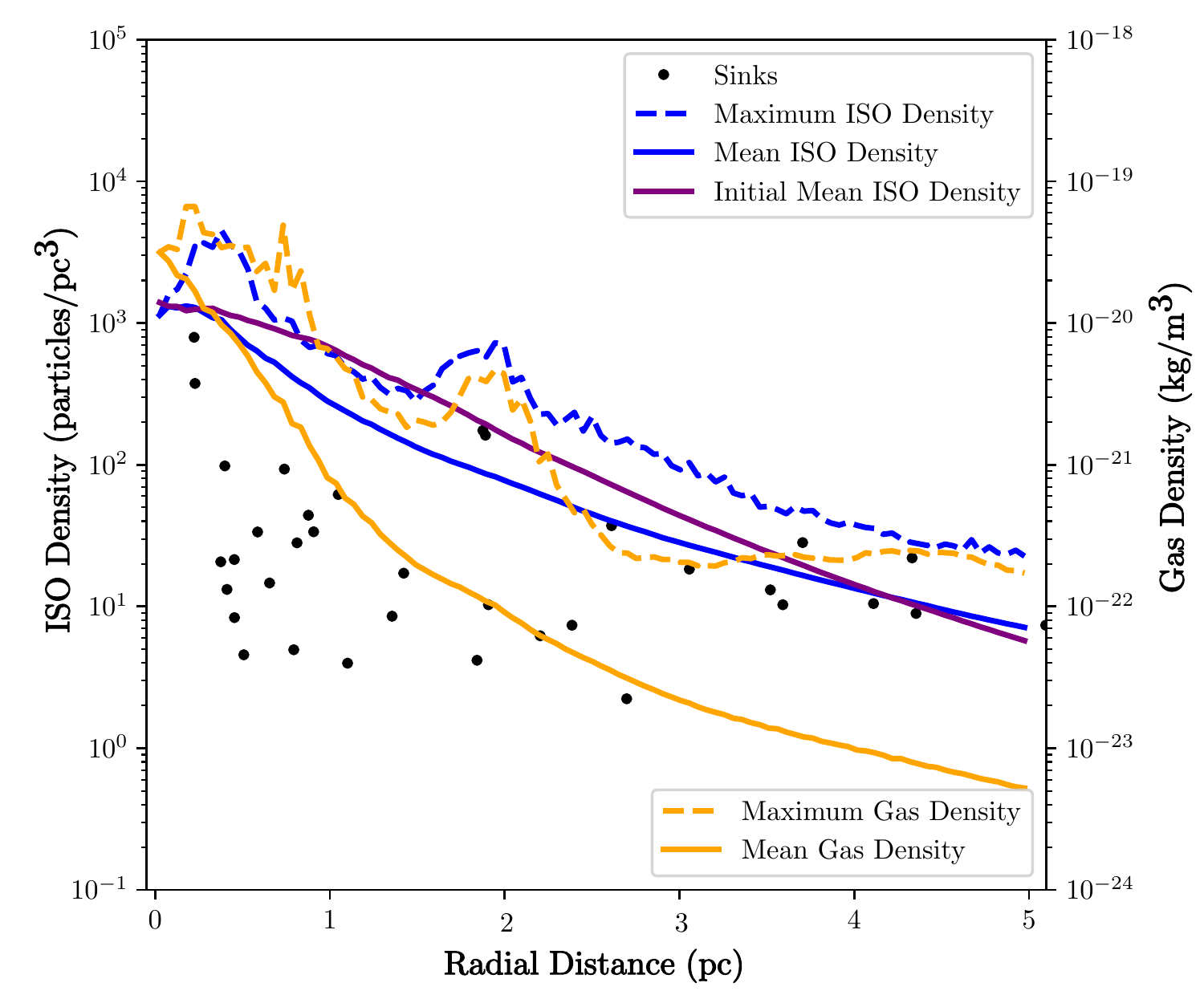}
\caption{
Radial profile snapshots of the evolution of the gas (orange) and interstellar object (blue) density relative to the cluster centre for the single Pl-30k molecular cloud collapse simulation. 
Black dots indicate the sink particles.
The density of interstellar objects (ISOs) is obtained using Eq. 2 and radially averaged. It peaks notably around certain sink particles, following them as their distance to the cluster centre varies throughout the simulation.
The full movies of this and the other three simulations are available at \doi{10.5281/zenodo.4301100}.}
\label{fig:snapshot}
\end{figure}

To develop a more quantitative understanding of the ISO dynamics, we first need to understand the stars' dynamics. 
Fig. \ref{fig:sinkproperties} shows the formation history (left) and mass evolution (right) of the sink particles in all four molecular cloud collapse simulations. 
In our higher-resolution setups, the first sinks form at $\sim$ 0.3 Myr, which corresponds to about 0.4 free-fall times (Fig. \ref{fig:sinkproperties}, top axis).  
In the lower-resolution simulations, sink formation tends to set in slightly later. 
An abundance of sinks form at these early stages. 
However, the number of sinks quickly decreases, due to many mergers and ejections: only about a quarter to a third of the initial number of sinks survive to three free-fall times. 
The mass distribution development indicates that the merging of sinks is the dominant growth process, as the total mass in sinks grows very little after the number of sinks saturates. 
In summary, initially, many sinks form; however, they rapidly merge within a single free-fall time. After this coagulation process, the systems become dominated by a few tens of massive sinks. 

The next question is whether the ISOs only generally follow the overall gas collapse, or do truly become gravitationally bound to individual sinks --- in which case the ISO density should be higher around the forming stars than elsewhere. 
We investigate\footnote{This can be best seen in the movies available at \doi{10.5281/zenodo.4301100}.} the respective densities of the gas and ISO particles as a function of radial distance from the simulation centre, together with the motion and mass of the sinks.
Crucially, across all four simulations, the density of ISOs is high in areas where the gas density is high and there are either many sinks or one or more high-mass sinks.
As a representative snapshot, Fig. \ref{fig:snapshot} shows the variation in the ISO population with an initial Plummer density distribution, for the Pl-30k simulation\footnote{This simulation's movie is named Pl-30k\_radial-density.mp4}. 
At time $t=0.34$ Myr, sink particles form at the highest gas densities. 
Soon, ISO concentration starts to peak close to the highest mass sink particles; this is most easy to spot for the binary forming at $\approx$ 2 pc. 
Note the local ISO density maxima at $\sim$ 0.4 pc and $\sim$ 1.7 pc in Fig. \ref{fig:snapshot}a, where the accumulation of many sinks attracts a large fraction of the ISOs. 
This effect also occurs when there is just one or two high-mass sinks, like the two massive sinks at 1.1 pc in Fig. \ref{fig:snapshot}b. 
These two sinks lead to a distinctive double peak in the ISO density --- and the ISO density peak follows this pair of sinks as the binary's distance to the cluster center varies throughout the simulation.
In general, density peaks follow sinks.
This means that not only do ISOs follow the overall collapse of the molecular cloud, leading to an average higher density of ISOs, but crucially, implies they become bound to individual sinks during the star formation process. 

The entire collapse process is very dynamic with sinks merging, but also leaving the simulation area as a result of scattering. The ISO population follows these escapees unless their velocity is very high --- in this case, the ISOs stay behind. 
This is the reason for the decrease in ISO density at later times: these ISOs have fully taken part in the collapse leading to sink formation.

\begin{figure}[t]
\includegraphics[width=0.6\textwidth]{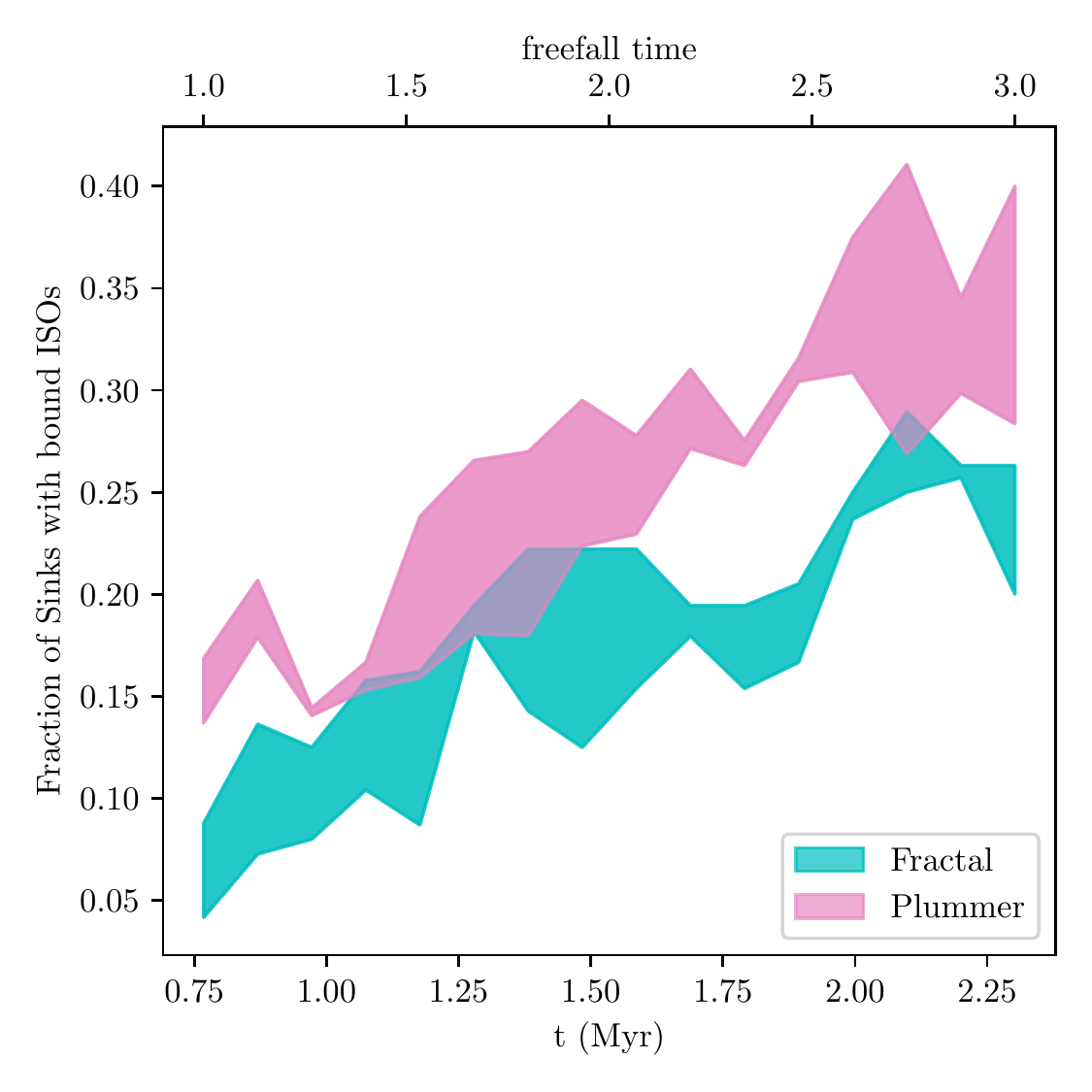}
\caption{
The fraction of the sink particle population that host a bound population of interstellar objects (i.e. are populated sinks), as a function of free-fall time throughout all four molecular cloud collapse simulations. 
Colour indicates the two initial gas distributions of the simulations: fractal (green) and Plummer (pink), while the two simulation resolutions of $3-6 \times 10^4$ gas particles create the variance boundaries.
}
\label{fig:ISO_free_fall}
\end{figure}

We now closely consider the ISO test particles that are gravitationally bound to individual sinks across all simulations. 
There are two conditions to be satisfied for an ISO to be considered bound: 
(i) the binding energy is negative and 
(ii) the contribution to the ISO's acceleration is largest from that sink (compared to the cumulative action of all other sinks).

Fig. \ref{fig:ISO_free_fall} shows the fraction of ISOs bound to sinks as a function of time. 
The maximum and minimum fraction of bound ISOs are shown separately for the Plummer and fractal initial distributions. 
As we saw in Fig. \ref{fig:sinkproperties}, the sink particle number becomes relatively stable only after 0.75 Myr of cloud evolution, therefore the bound fraction is only shown from 0.75 Myr onwards in Fig. \ref{fig:ISO_free_fall}.
Populated sinks steadily increase overall through the cloud collapse, regardless of how the gas of the molecular cloud was initially distributed: the general trend is a roughly linear temporal increase of the fraction of ISOs bound to sinks. 
The symmetric Plummer initial ISO distribution leads to a consistently higher fraction of populated sinks, $\approx$ 35\%,  than a fractal distribution, $\approx$  25\%. 
The difference in the percentage of sinks capturing ISOs begins early on and continues throughout the cloud collapse. 
A likely explanation is that the steady gradient in a Plummer potential allows for more efficient binding of ISOs than do strongly locally varying fractal ones. 

The low resolution of these simulations mean their bound populations of ISOs are only lower limits on the intrinsic populations present at the nascent star systems.
Capturing one test particle is equivalent to a sink seizing at least 10$^{10}$ to 10$^{14}$ real ISOs.
Thus a large fraction of sinks capture considerable amounts of ISOs: a quarter to a third of sinks capture at least 10$^{10}$ to 10$^{14}$ real ISOs.
Sinks that capture a lower number of ISOs are below the resolution limit and not yet quantified.

\begin{figure}[t]
\includegraphics[width=0.99\textwidth]{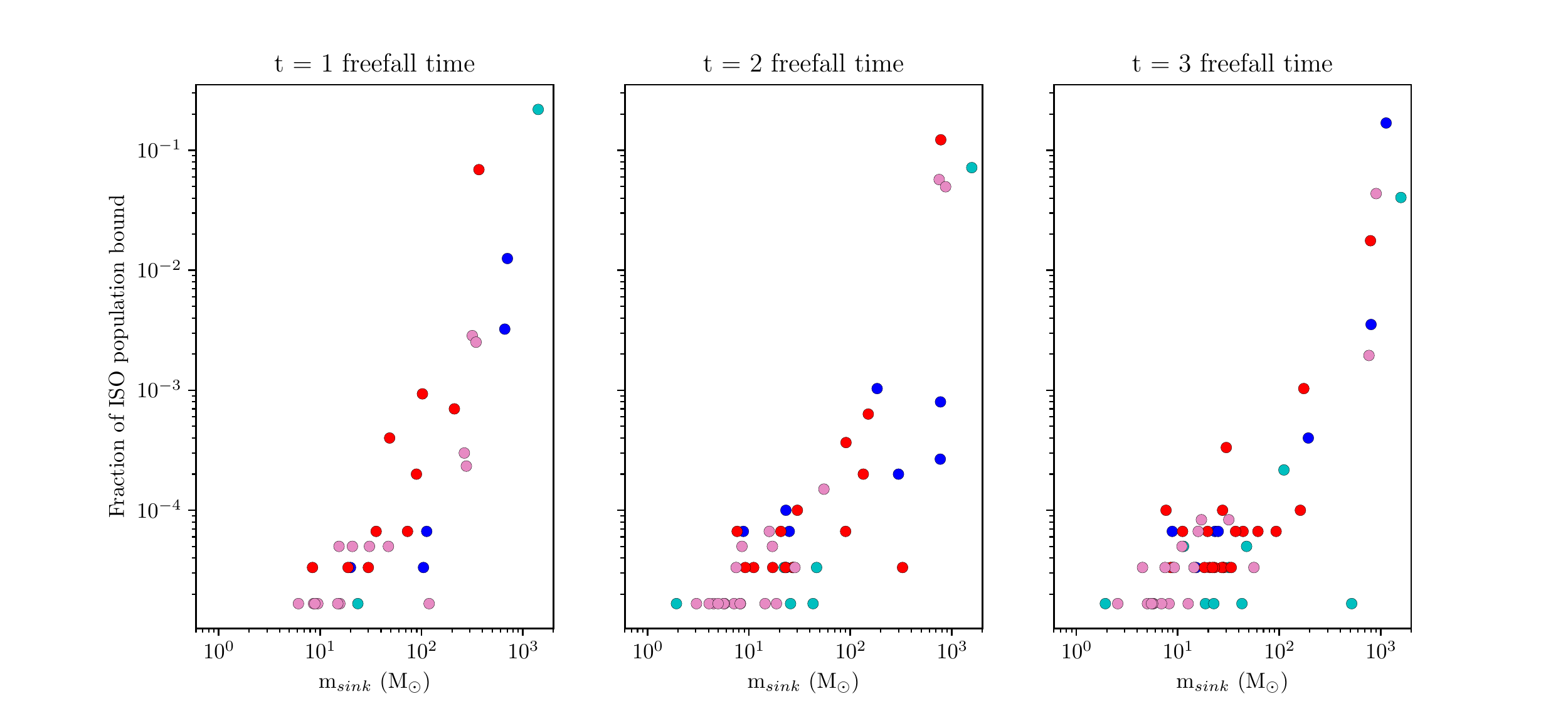}
\caption{
The fraction of the interstellar object population that are bound to sink particles of varying mass, for three snapshots of free-fall time throughout all four molecular cloud collapse simulations. 
As in Fig. \ref{fig:sinkproperties}, point colours indicate the four simulations' parameters of gas initial conditions and resolution. 
}
\label{fig:sink_mass}
\end{figure}

These simulations illustrate that there are large variations in how efficient sinks are in capturing ISOs: key parameters are sink mass and velocity.
The higher a sink's mass, the more ISOs it captures (Fig. \ref{fig:sink_mass}). 
Such a behavior could be expected from the gravitational force law; the proportion of ISOs bound to a sink particle should scale with the sink mass. 
Indeed, this is approximately valid for the sinks with masses in the range $ 5\MSun\ < M_s < 100 \MSun$. In this sink mass range, the approximate relation for $\epsilon_{\mbox{iso}}$, the number of ISOs, holds 
\be
\epsilon_{\mbox{iso}} \approx 10^{-5.5 \pm 0.5} M_{\mbox{sink}},\label{eq:mass}
\ee
where $M_{\mbox{sink}}$ is the mass of the sink particle. 
However, it seems that very high-mass sinks can often capture considerably more ISOs than expected from this relation.  
A single high-mass sink ($>$ 200 \MSun\ ) often captures 20 or even 30 percent of all ISOs in the simulation. 
A possible reason for this hyper-capturing is that the high mass sinks are formed through merger events. 
Here, the Hill spheres of the individual sinks are merging. 
As the Hill sphere scales as $M^{3/2}$, merging sinks capture more ISOs due to their larger Hill sphere. 
Through this merger, already-bound ISOs will become more strongly bound, while the nearby ISOs feel a more substantial potential and are captured.  
From star formation, the process of competitive accretion  of gas to high-mass stars is known \citep{Bate:2003}; it seems a similar effect occurs for ISOs --- competitive capture to high-mass sinks. 
It also resembles oligarchic growth in planetesimal formation.

\begin{figure}[t]
\includegraphics[width=0.7\textwidth]{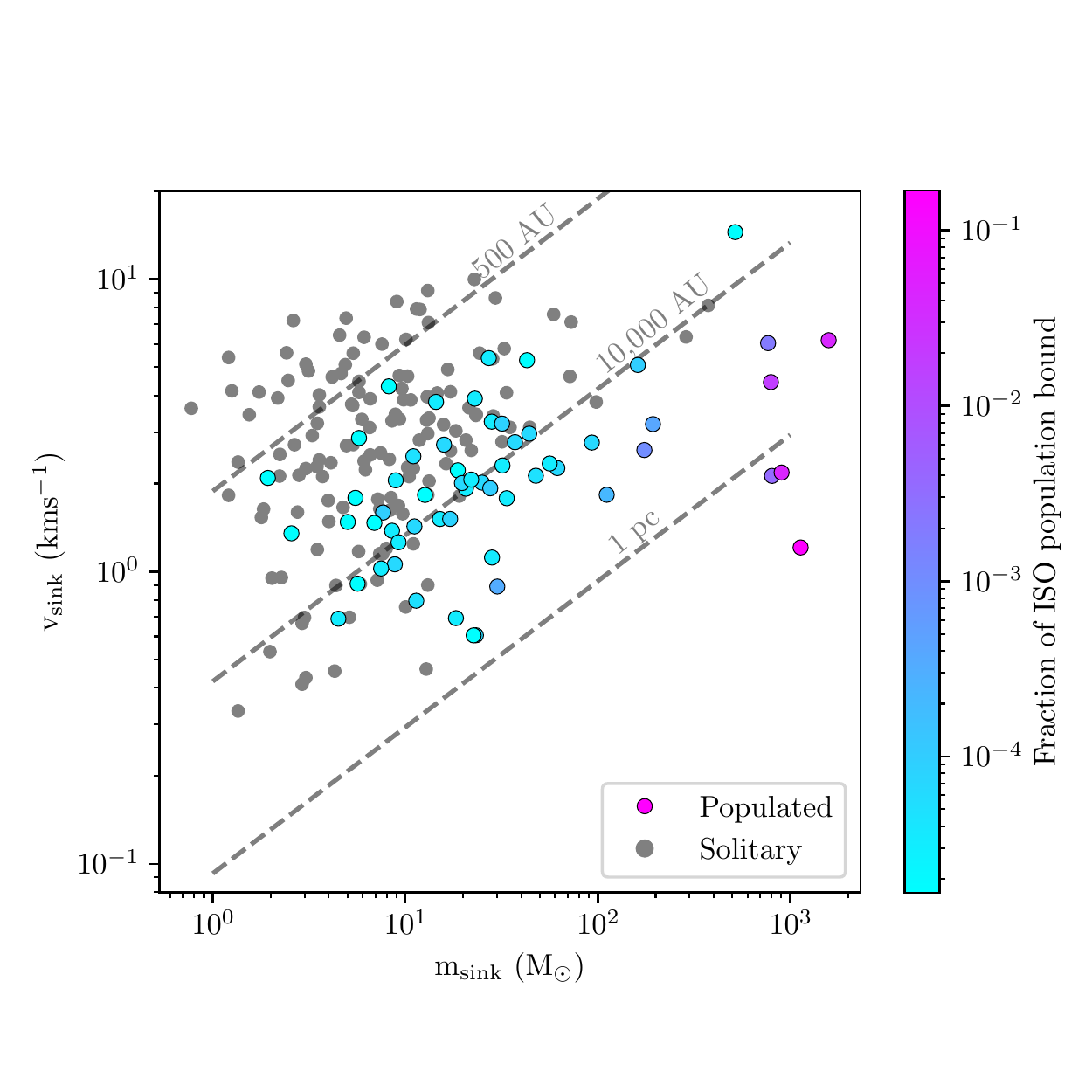}
\caption{
The relative interstellar object populations of all sinks, compared with the sink particle velocity as a function of sink mass, from all simulations after three free-fall times. 
Grey circles indicate ``solitary" sinks that have no ISOs bound to them, whereas coloured circles show ``populated" sinks with an ISO population; colour indicates the fraction of a simulation's ISOs that an individual sink binds to itself.}
\label{fig:Vel}
\end{figure}

Sink velocity is the other key parameter that determines a sink's ability to attract and retain ISOs. Fig. \ref{fig:Vel} shows the velocities of all the sink particles from the various simulations vs their masses. 
Sinks that do not bind any ISO test particles to themselves are indicated in grey; sinks with bound ISOs are colour-coded according to the fraction of ISOs they capture. 
The sinks' velocity plays a significant role when it comes to binding ISOs to them.
The majority of solitary sinks are speedy low-mass particles, while slow, more massive sinks tend to retain their ISO populations across a range of velocities. 
A typical example of high-velocity sinks are the ones ejected from the forming cluster; for these high-velocity sinks, two processes are at work. 
First, at such high velocities, typically above \mbox{5 km/s,} the ISOs cannot follow sinks.  
Second, close encounters between sinks, which lead to ejections of the less-massive sinks, can also redistribute the ISOs between the interacting sinks: more massive sinks can capture ISOs from less massive ones. 
However, as we see even some massive and high-velocity sinks losing their ISOs, it appears that the first process dominates.  

The gravitational binding to massive sinks is stronger; therefore, ISOs remain bound to massive sinks that are travelling at higher velocities than they do for low-mass sinks. 
Thus the actual value of maximum velocity $v_{max}$ where ISOs can still follow is also a function of the cluster mass.  
In our simulations this can be approximated by $v_{max} \approx 2\ \mathrm{km/s} \times (M/\MSun)^{1/3}$. 
However, this approximation relies heavily on the single data point in the top right corner of the figure.
Any sink with a velocity $v_s > v_{max}$ is travelling too fast to capture any additional ISOs, and potentially, already-bound ISOs can no longer follow the sinks. 
A consequence of the mass-dependence of $v_{max}$ is a robust natural gradient in the bound ISO population's size across the mass-velocity range of sink particles. 
High-mass sinks, the numerical representatives that would form into small star clusters, can still keep at least some of their ISOs bound, even if they move at speeds in excess of 20 km/s.
However, at a given sink mass, in general slow sinks keep a much larger portion of their ISOs bound to them than fast ones.

\begin{figure}[t]
\includegraphics[width=0.8\textwidth]{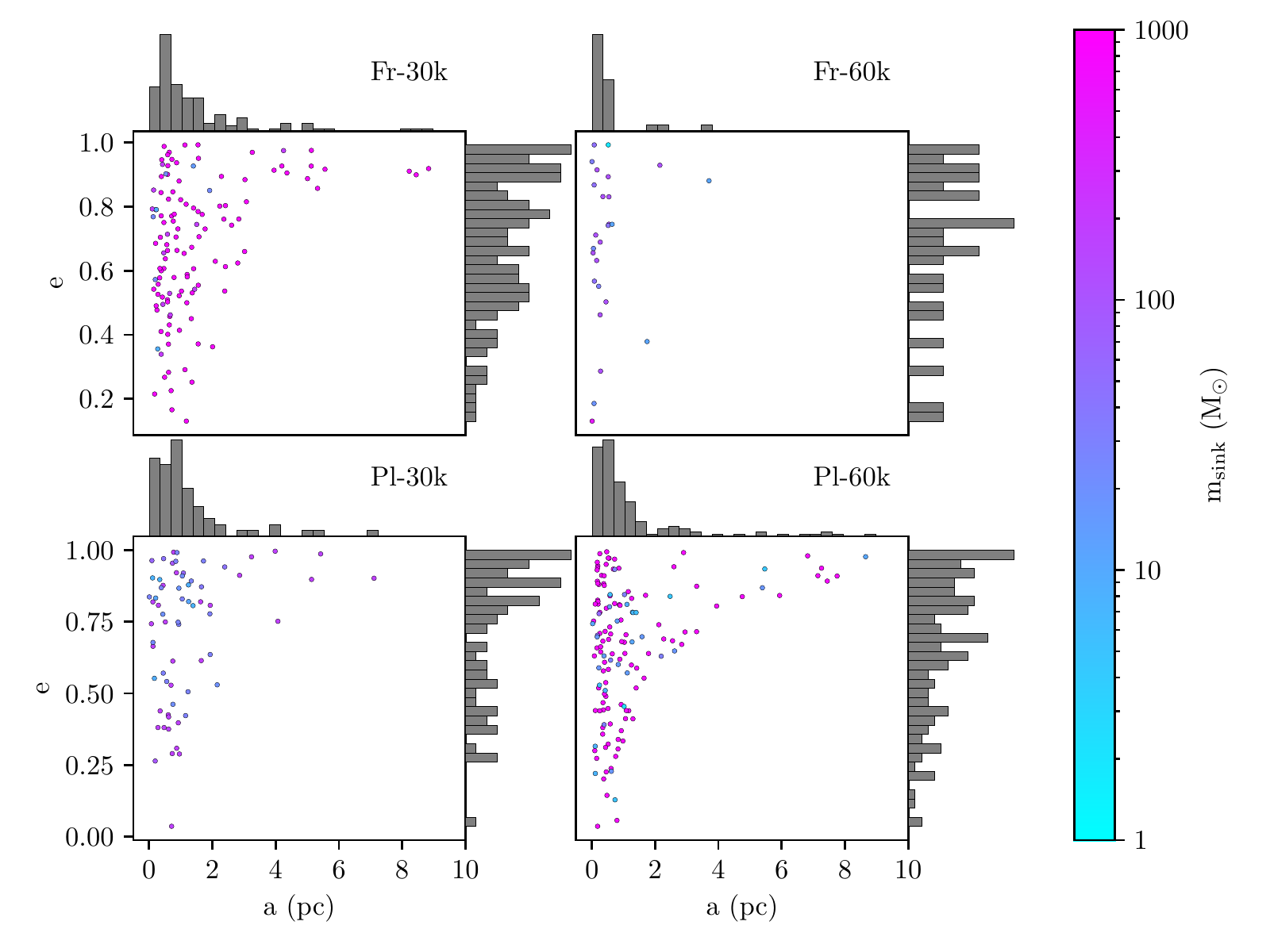}
\\
\includegraphics[width=0.8\textwidth]{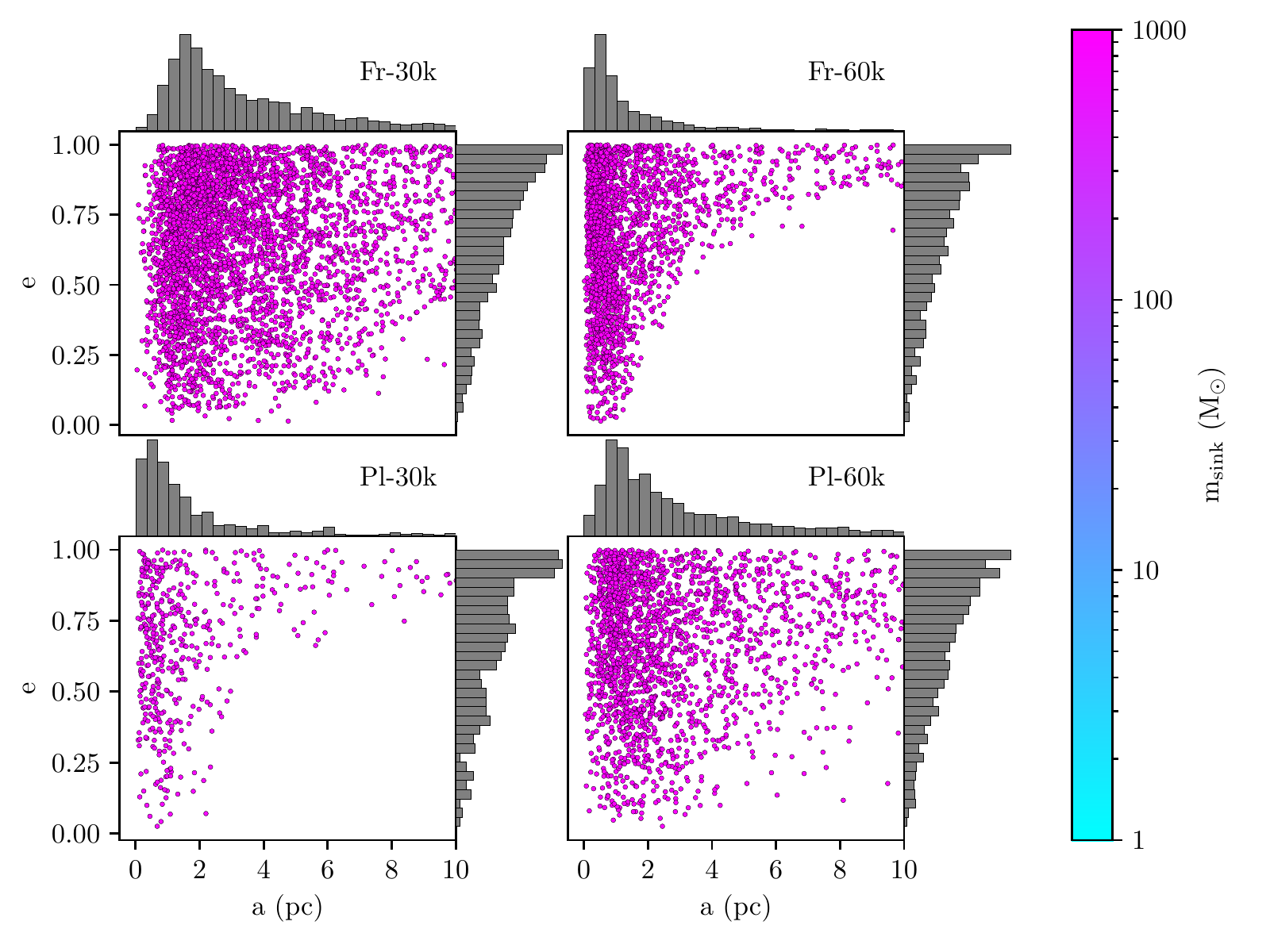}
\caption{
The orbital element distribution of the bound interstellar objects of populated sinks in each of the four cloud-collapse simulations, after three free-fall times.
\textit{Top:} semi-major axis and eccentricity distribution of ISOs bound to all populated sinks except the most massive. \textit{Bottom:} semi-major axis and eccentricity distribution of ISOs bound to the most massive populated sink.
Simulation parameter labelling is as in Fig.~\ref{fig:sinkproperties}, and ISO point colour indicates its sink's mass.
}
\label{fig:peri_ecc}
\end{figure}

Finally, we consider the orbital parameters of the ISO particles bound to sinks across all simulations. 
Fig. \ref{fig:peri_ecc} shows a scatter plot of the semi-major axes and eccentricities of the orbits of the ISOs bound to a sink.  
To clarify the role of the most massive sinks, which clearly retain a dominant fraction of the bound population, the bottom panel displays the parameters for all ISOs bound to the most massive populated sink, whereas the top panel illustrates the same parameters for all other ISOs bound to sinks, with the exception of this most massive sink. 
In all simulations, most ISOs move on orbits with a semi-major axis below 1 pc, with many orbits being highly eccentric. 
The median semi-major axis of the ISOs bound to high-mass sinks is 2.3\,pc compared to 0.75\,pc of the lower mass sinks. 
The semi-major axes and eccentricities translate into median periastron distances of 0.56\,pc and 0.20\,pc, respectively.

Overall, bound ISOs retain the moderate to extreme eccentricities reasonable for orbits in highly perturbed systems that have not yet had time to circularise.
One question beyond the scope of the current study is whether the highly eccentric ISO orbits become circularized with time  --- in analogy to the circularising of eccentric planetary orbits in gas-rich discs \citep{Kikuchi:2014}. 
As the gas is present in the molecular cloud's collapse for at most 10--20 Myr, it remains unclear whether this period suffices for serious circularisation to take place.

Visually, the semi-major axis and eccentricity distributions of the bound ISOs in Fig. \ref{fig:peri_ecc} for all but the most massive sinks (top) appear similar to the corresponding distributions for the ISOs bound to the most massive sink (bottom). 
To test if the two samples (ISOs at less massive vs. most massive sinks) for a single simulation could have been drawn from a common distribution, we performed a Kolmogorov-Smirnov (KS) test. 
Each orbital parameter was tested independently. 
The $p-$values from the KS test are shown in Table \ref{tab:kstest}.
We can reject the null hypothesis that the semi-major axis values were drawn from a common distribution for the Fr-30k, Fr-60k and Pl-60k simulations. 
For all simulations, we cannot reject the null hypothesis that the eccentricities in the two samples were drawn from a common distribution.

We also check if the samples of the orbital parameters of the ISOs bound to the largest sinks from each different simulation could have been drawn from the same distribution. 
We performed a two-sample KS test on the most massive sink ISO orbital parameter samples for each simulation pair, eg. Fr-30k with Fr-60k, Fr-30k with Pl-30k etc. 
For every pair of simulations, we can reject the null hypothesis that the semi-major axis samples were drawn from a common distribution, where the maximum $p-$value was $5.2\times 10 ^{-5}$. 
For every pair of simulations, we cannot reject the null hypothesis that the eccentricity samples were drawn from a common distribution, where the minimum $p-$value was $0.18$. 
The above statistical tests demonstrate that within each simulation, the distribution of semi-major axis values among the bound ISOs depends on the individual sink properties. 
The eccentricities of the bound ISO populations, however, are likely to be common across all simulations and are largely independent of the sink properties.

\begin{table}[]
\caption{$p-$values for the two-sample Kolmogorov-Smirnov test for the samples in Fig. \ref{fig:peri_ecc}. }\label{tab:kstest}
\begin{center}
\begin{tabular}{|c|c|c|c|c|}
\hline
  & Fr-30k   & Fr-60k  & Pl-30k & Pl-60k   \\ \hline
a & $2.54\times 10^{-31}$ & $3.86\times 10 ^{-7}$ & 0.365  & $3.33\times 10^{-16}$ \\ \hline
e & 0.693    & 0.985   & 0.242  & 0.711    \\ \hline
\end{tabular}
\end{center}
\end{table}

\section{Consequences}

We found that the most massive emerging sinks bind disproportionately many ISOs to themselves.  
As the most massive sinks represent subclusters of stars, this could mean that stars that are born in subclusters have a higher than average ISO population bound around them. 
One possible consequence could be a larger or more densely populated Oort cloud for the resulting systems.

Considering the solar-mass scale of our sinks instead, there are potential implications for the Solar System.
Extrapolating Eq.~\ref{eq:mass} down in mass to solar-mass sink particles, and applying a size distribution relevant for comets such as \cite{Hughes:1990} to Eq.~\ref{eq:mass}'s population constraint for massless ISO test particles, approximately 3 Earth masses of ISOs would be present at solar-mass sinks.
This has a pleasing coincidence with plausible Oort cloud masses --- as uncertain as those remain \citep{Dones:2015}.
It could imply that the order-of-magnitude scale of the Solar Oort cloud is rather typical of the number of ISOs that a solar-type star can capture in its natal state.
Such events contrast with later dynamical mechanisms to obtain comets from Solar birth cluster stellar siblings \citep[e.g.][]{Levison:2010, Belbruno:2012} or direct capture from the field ISO population \citep[e.g.][]{Napier:2021}.
In the absence of full cloud-to-migration simulations of Oort cloud formation, this most likely represents a massive overestimate of the current ISO fraction present in the Solar System's Oort cloud. 
However, sink binding is a novel dynamical mechanism for the current Oort cloud to contain some remnant, entirely pre-solar ISOs. 
While all Oort cloud material was at one point in the ISM, leading to a complex, processing-dependent synergy between interstellar and cometary chemistry \citep{Mumma:2011}, the distinctive high CO abundance of 2I shows that a truly interstellar comet can retain an unusual fingerprint  relative to the currently known comet populations \citep{Bodewits:2020,Cordiner:2020}. 
As Oort cloud comets ever continue to fall into the bounds of observability, future population predictions will be testable.

While the size distribution of interstellar objects is presently insufficiently observationally constrained to debate, it is reasonable to expect that it will range from presolar grains through sizes and masses similar to \Ou\ and 2I/Borisov and upward to dwarf planets.
Indeed, free-floating planets could be regarded as the very high-mass end of the ISO population \citep{Mroz:2020}. Consequently, sinks will capture a vast majority of small icy or rocky bodies --- but in a few rare cases, free-floating dwarf planets, or even planets.  
A scenario to consider is thus the existence of a stellar/planetary system with a previous-generation planet in a widely separated orbit --- a lower-mass companion than those currently seen for brown dwarfs.
Additionally, in an extremely rare event, some stars might be born with previous-generation planets orbiting them. 
The presence of such ``ancient" planets might have far-reaching consequences for the disks forming around these stars and the planetary systems eventually emerging from them.

When ISOs bind around the forming stars, ISOs are depleted elsewhere in the interstellar medium. 
Consequently, the concentration of ISOs in MCs would ultimately mean that MCs deplete the ISO population in their immediate galactic environment. 
As MCs form and dissolve throughout the Milky Way, this could result in a rather inhomogeneous ISO concentration in the ISM, on timescales of several tens of Myr. 
This mechanism is independent of requiring substantial populations of low-area-density ISOs, in contrast to the process to concentrate ISOs in stellar streams proposed by \citet{Eubanks:2019}.
With inhomogeneity in play, the question arises of how typical the local ISO concentration that we measure with surveys in the Solar System is for the entire Galaxy (see also \citet{Pfalzner:2020}).

\section{Limitations}
\label{sec:discussion}

With this proof-of-principle study, we demonstrate that ISOs not only follow the gas's collapse, but actually bind to individual sinks when a molecular cloud starts to collapse and form stars. 
We deliberately opted for low-resolution simulations to obtain a first impression of whether ISOs react at all to the collapse of a cloud. We consider the limitations of this qualitative study, together with some initial implications. However, first we want to provide guidelines for future simulations that can give more quantitative results.

Unquestionably, in future studies, the simulations' resolution will have to be increased considerably; this is valid for the gas and the ISOs alike.
The low resolution manifests itself in two main areas: (i) the sink particle masses being relatively high, and (ii) each of our ISO test particles representing the substantial population of 10$^{10}$--10$^{14}$ ISOs. 
This limits the reliability of the quantitative results in this study: Eq. \ref{eq:mass} and the maximum sink velocity at which ISOs can still remain bound to a sink require further testing. 

A limiting consequence of the low resolution in the gas is that sinks often do not form a single star, but an entire small cluster of stars. 
Future higher simulation resolutions should allow for a one-to-one representation between sink particles and subsequent stars. 
Including an improved description of feedback should be a prime target, together with a more realistic star formation efficiency.
This better match to the stellar initial mass function will permit exploration of whether there are differences in the capture of ISOs by single stars compared to that by binary systems.

The low resolution of the ISOs simulated here means that only events that correspond to extremely high capture rates of ISOs are recorded. 
Any event that leads to the capture of less than 10$^{10}$ ISOs is not quantified. 
This means, for example, that we can just barely resolve the capture of a population equivalent to that of the Oort cloud ($\lesssim 10^{11}$ comets; \citet{Dones:2015}) in the current simulations. 
As one outcome of this study is that ISO capture rates vary considerably from sink to sink, we likely miss all the events when ``just" a few billion ISOs are captured rather than trillions. 
Better resolution of smaller ISO populations, together with resolving the individual stars forming from these high-mass sinks, will allow the sink mass-dependence of ISO capturing to be fully understood. 
Currently it is unclear how the neglected presence of ISOs influences star and planet formation --- it is yet to be established what minimum number of ISOs would be needed to lead to a significant impact in the consecutive processes. 

Future simulations should address the influence of the initial spatial and velocity distributions of the ISO population. 
Here we adopted a fractal and a concentrated spatial distribution.  
However, if ISO production is homogenised within the ISM and therefore does not follow the cloud formation process, a homogeneous spatial distribution with some small random noise would be a realistic option. 
Equally, a variety of velocity distributions should be tested, as the dominance of the various ISO production processes and therefore ejection speeds from the origin systems are yet to be fully characterised in ensemble \citep[e.g.][]{Rice:2019,Pfalzner:2021}.

We have demonstrated that quantitative studies of ISO capture during collapse will require (i) a much higher resolution and (ii) detailed inclusion of more physical effects. 
The scientific question dictates the required resolution because it determines the relevant spatial scales.  
The fundamental questions next to be addressed in the context of planetary system formation are whether ISOs only become loosely bound to the stars, or if they are incorporated in large numbers in the disks forming around the stars and are accreted. 
In the latter, the fundamental scales are the disk size and the accretion area: modern star formation simulations can achieve within the range of the sizes of disks, but resolving the accretion area in collapse simulations seems to be out of reach at the moment due to limitations in computing power. 
However, increasing the simulation resolution by a factor of ten from our work here will not result in a sizable advance in knowledge. 
In order to resolve even on a scale of 10,000 au, which would cover the inner edge of the current Solar Oort cloud, but not a typical 100 au-sized planet-forming disk, one would need to resolve the system at a density scale of $\sim 10^{-19}$~g/cm$^3$, which requires an increase in the number of gas particles by a factor of ${\cal O}(10^4)$, increasing the gas-particle resolution to $10^9$ particles, or 5\,M$_\oplus$ per gas particles. 
To resolve a cloud of bound ISOs down to the level of $10^4$ ISOs per particle would require $10^8$ ISO particles in our calculations.
%The number of ISO test particles should be approximately of the same order as only then it becomes feasible to determine the incorporation into the disc. 
Both resolutions, in gas density as well as in ISOs, are in principle feasible, while beyond the scope of the current work. 
We hope that this study motivates other researchers who focus on the star formation process to include ISOs in their future studies.

\section{Summary}
\label{sec:conclusion}

There are strong indications that ISOs are omnipresent in large numbers throughout the interstellar medium. 
Thus, ISOs should be considered as natural an ingredient to star formation as are gas and dust. 
However, the role of ISOs in the star formation process is currently unclear. 
Here we present a study that includes ISOs in simulations of the star cluster formation process.  
These simulations have to be regarded as a proof-of-principle investigation, as we deliberately treat the star formation process itself in a very simplified way.
Nevertheless, we lay the foundations for future, more detailed studies. 
This first attempt at tackling the question of the fate of ISOs during cloud collapse already reveals the following general trends:

\begin{itemize}
\item This work shows that \emph{ISOs take part in the cloud collapse}, despite being decoupled from the gas hydrodynamics.  As the ISOs follow the cloud collapse, the ISO concentration close to sinks increases.
\item The degree of ISO concentration depends on the collapse conditions. We find that spherically-concentrated potentials (Plummer potential) favor ISO capture compared to fractal potentials. In our simulations, after three free-fall times, typically 35\% of sinks have captured ISOs in the case of Plummer profiles compared to 25\% for fractal distributions. Both values are lower limits, due to the relatively low resolution of our simulations.
\item Massive sinks bind considerably more ISOs than low-mass sinks (see Eq. \ref{eq:mass}) due to a competitive ISO capturing process.  Extrapolating to a typical solar-mass star, we find that it binds the equivalent of several Earth-masses of material in ISOs. This value is comparable to the current mass of our Oort cloud.
\item Over the first three free-fall times, the fraction of sinks with ISOs bound to them increases  linearly with time. One of the reasons is that sinks grow by merging. However, more importantly, sinks attract additional ISOs as they move through the collapsing molecular cloud. As a result, massive sinks tend to have a disproportionate late-stage fraction of an initial population of ISOs.
\item Once captured, ISOs usually remain bound to their sinks. However, if sinks are ejected by encounters and move relatively fast, ISOs either do not become bound in the first place or become unbound later as they cannot follow the fast-moving sinks.  The actual maximum velocity at which ISOs can remain bound to the sinks depends on the sink mass; however, a typical value is 5 km/s.
\item Most ISOs are captured on relatively wide eccentric orbits; in these molecular cloud conditions, a median periastron distance of 0.56\,pc and a mean eccentricity of 0.75.  It will require further studies to determine on what timescales these ISOs move onto closer orbits and become circularized due to their interaction with the gas.
\end{itemize}

The concentration of ISOs in molecular cloud collapse and the capture of ISOs by the sinks is a first indication that there exists a \emph{life cycle of ISOs}, similar to that found for gas and dust \citep{Pfalzner:2019, Grishin:2019}. 
The ISO life cycle starts with the formation of planetesimals in the parent system and their ejection into the interstellar medium. 
It continues by incorporating ISOs into molecular clouds, where the ISOs follow the cloud collapse.
Future simulations will test if the ISOs eventually take part in the next generation of stars and the planetary system evolving around them.

One surprising outcome of this investigation is that the fraction of the initial ISO population in the molecular cloud that becomes bound can differ considerably between individual sinks. 
Sinks may be entirely solitary, or orbited by a rich fraction of a cloud's initial ISO population.
It has been proposed that the presence of ISOs could accelerate planet formation in a new system \citep{Pfalzner:2019}. 
It will be fascinating to see whether the system's formational abundance of ISOs can link to the remarkable observed variety in planetary systems and timescales for planet formation. 

\section*{Energy consumption of these calculation}

The calculations using the hydrodynamics solver 
are performed on the Little Green machine II and took about 
$10^5$ CPU hours. The calculations were performed on 48 cores (4 nodes with 12 cores) of 2.4GHz consuming  190Watt per node.  We then arrive at a total power consumption of about 200KWh. We estimate our computation produced about
50kg CO$_{2}$ \citep{2020NatAs...4..819P}, which is enough to drive a Fatboy Special motorbike from Leiden to J\"ulich, and back.

\acknowledgments
We thank the referee, who has provided very helpful comments that have considerably improved the manuscript.

\bibliographystyle{aasjournal}
\bibliography{references}

\end{document}